\documentclass[]{aa}

\usepackage[dvipsnames]{xcolor}
\usepackage{float}

\def\rmit#1{{\it #1}}              
\def\specchar#1{{\sc #1}}

\def\SiI{\mbox{Si\,\specchar{i}}}

\def\CaII{\mbox{Ca\,\specchar{ii}}} 
\def\CaIII{\mbox{Ca\,\specchar{iii}}}

\def\eg{\rmit{e.g.}}

\usepackage{booktabs}
\usepackage{graphicx}
\usepackage{multirow}
\usepackage[normalem]{ulem}
\usepackage{color}
\usepackage[flushleft]{threeparttable}

\usepackage{array}
\newcolumntype{?}{@{\vrule width 2pt}}

\usepackage{hyperref}
\hypersetup{
    final=true,
    pageanchor=true,
    colorlinks=true,
    breaklinks=true,
    linkcolor=blue,
    citecolor=blue,
    urlcolor=blue,
    pdfpagemode=UseNone,
    pdftitle={},
    pdfauthor={T. Felipe et al.},
    pdfsubject={Solar Physics},
    pdfkeywords={Methods: observational -- Sun: chromosphere -- Sun: oscillations  -- sunspots -- Techniques: polarimetric}}

\titlerunning{Downflowing umbral flashes}   

\begin{document}



\title{Downflowing umbral flashes as an evidence of standing waves in sunspot umbrae}

\author{T. Felipe\inst{\ref{inst1},\ref{inst2}}
\and V. M. J. Henriques\inst{\ref{inst3},\ref{inst4}}
\and 
J. de la Cruz Rodr\'{\i}guez\inst{\ref{inst5}}
\and 
H. Socas-Navarro\inst{\ref{inst1},\ref{inst2}}
}


\institute{Instituto de Astrof\'{\i}sica de Canarias, 38205, C/ V\'{\i}a L{\'a}ctea, s/n, La Laguna, Tenerife, Spain\label{inst1}
\and 
Departamento de Astrof\'{\i}sica, Universidad de La Laguna, 38205, La Laguna, Tenerife, Spain\label{inst2} 
\and
Institute of Theoretical Astrophysics, University of Oslo, P.O. Box 1029 Blindern, NO-0315 Oslo, Norway\label{inst3} 
\and
Rosseland Centre for Solar Physics, University of Oslo, P.O. Box 1029 Blindern, NO-0315 Oslo, Norway\label{inst4} 
\and 
Institute for Solar Physics, Dept. of Astronomy, Stockholm University, AlbaNova University Centre, 10691, Stockholm, Sweden\label{inst5}
}

\abstract
{Umbral flashes are sudden brightenings commonly visible in the core of some chromospheric lines. Theoretical and numerical modeling suggest that they are produced by the propagation of shock waves. According to these models and early observations, umbral flashes are associated with upflows. However, recent studies have reported umbral flashes in downflowing atmospheres.} 
{We aim to understand the origin of downflowing umbral flashes. We explore how the existence of standing waves in the umbral chromosphere impacts the generation of flashed profiles.}
{We performed numerical simulations of wave propagation in a sunspot umbra with the code MANCHA. The Stokes profiles of the \CaII\ 8542 \AA\ line were synthesized with the
NICOLE code.}
{For freely-propagating waves, the chromospheric temperature enhancements of the oscillations are in phase with velocity upflows. In this case, the intensity core of the \CaII\ 8542 \AA\ atmosphere is heated during the upflowing stage of the oscillation.  If we consider a different scenario with a resonant cavity, the wave reflections at the sharp temperature gradient of the transition region lead to standing oscillations. In this situation, temperature fluctuations are shifted backward and temperature enhancements partially coincide with the downflowing stage of the oscillation. In umbral flash events produced by standing oscillations, the reversal of the emission feature is produced when the oscillation is downflowing. The chromospheric temperature keeps increasing while the atmosphere is changing from a downflow to an upflow. During the appearance of flashed \CaII\ 8542 \AA\ cores, the atmosphere is upflowing most of the time, and only 38\% of the flashed profiles are associated with downflows. }
{We find a scenario that remarkably explains the recent empirical findings of downflowing umbral flashes as a natural consequence of the presence of standing oscillations above sunspot umbrae.}


\maketitle


\section{Introduction}

Oscillations are a ubiquitous phenomenon in the Sun. Their signatures have been detected almost everywhere, from the solar interior to the outer atmosphere. One of the most spectacular manifestations of solar oscillations are umbral flashes (UFs). They were discovered by \citet{Beckers+Tallant1969} \citep[see also][]{wittmann1969}, who observed them in sunspot umbrae as periodic brightness enhancements in the core of some chromospheric spectral lines. Since their first detection, UFs were associated with magneto-acoustic wave propagation in umbral atmospheres \citep{Beckers+Tallant1969} and soon later \cite{Havnes1970} suggested that UFs in \CaII\ lines were formed due to the temperature enhancement during the compressional stage of these waves.     

The analysis of spectropolarimetric data has become a common approach for the study of UFs during the last two decades, since the studies undertaken by \citet{SocasNavarro+etal2000a, SocasNavarro+etal2000b}. They found anomalous polarization in the Stokes profiles and interpreted it as indirect evidence of very fine structure in UFs, unresolved in arc-sec resolution observations. Their model had one hot upflowing component embedded in a cool downflowing medium. Further independent, but still indirect, evidence of such fine structure was obtained by \citet{Centeno+etal2005}. Subsequent studies, benefiting from observations acquired with higher spatio-temporal resolution, were able to resolve those components in imaging observations \citep{SocasNavarro+etal2009,Henriques+Kiselman2013,Bharti+etal2013, Yurchyshyn+etal2014, Henriques+etal2015} and in full spectropolarimetric observations \citep{RouppevanderVoort+delaCruzRodriguez2013, delaCruz-Rodriguez+etal2013,Nelson+etal2017}. Non-local thermodynamical equilibrium (NLTE) inversions of \CaII\ 8542 \AA\ sunspot observations using the NICOLE code \citep{SocasNavarro+etal2015} have found that UF atmospheres are associated with temperature increments of around 1000~K and strong upflows \citep{delaCruz-Rodriguez+etal2013}. These observational measurements exhibit a good agreement with the theoretical picture of UFs as a manifestation of upward propagating slow magneto-acoustic waves. The amplitude of these waves increases with height as a result of the density stratification, leading to the development of chromospheric shocks \citep{Centeno+etal2006, Felipe+etal2010b} and UFs \citep{Bard+Carlsson2010, Felipe+etal2014b}. The properties of those UF shocks have been recently evaluated by interpreting spectropolarimetric observations with the support of the Rankine-Hugoniot relations \citep{Anan+etal2019, Houston+etal2020}.

The first NLTE inversions of UFs \citep{SocasNavarro+etal2001} were computationally very expensive (typically several hours per individual spectrum) and therefore only a few profiles from the same location and different times were inverted. Later improvements in code optimization, parallelization, and computer hardware allowed \citet{delaCruz-Rodriguez+etal2013} to analyze a few 2D maps with spatial resolution. As this trend continued, it is now possible to make UF studies with higher spatial and temporal resolution and much better statistics \citep[\eg,][]{Henriques+etal2017, Joshi+delaCruzRodriguez2018, Houston+etal2018, Houston+etal2020}. These recent results have revealed the existence of UFs whose Stokes profiles are better reproduced with atmospheres dominated by downflows \citep{Henriques+etal2017, Bose+etal2019, Houston+etal2020}.

Accommodating the downflowing UFs to the widely accepted scenario of upward propagating shock waves is a theoretical challenge. Studies employing sophisticated synthesis tools and numerical simulations of the umbral atmosphere have validated the upflowing UF model \citep{Bard+Carlsson2010, Felipe+etal2014b, Felipe+etal2018a, Felipe+EstebanPozuelo2019}. In those simulations, magneto-acoustic waves could propagate from the photosphere to the chromosphere. According to the phase relations of these propagating waves, velocity and temperature oscillations exhibit opposite phases (considering that positive velocities correspond to downflows), and the temperature increase causing the flash to match an upflow. However, recent studies have shown evidences of a resonant cavity above sunspot umbrae \citep[see][]{Jess+etal2020a, Felipe2020, Jess+etal2020b,Felipe+etal2020}. This cavity is produced by wave reflections at the strong temperature gradient in the transition region \citep[\eg,][]{Zhugzhda+Locans1981, Zhugzhda2008}. Waves trapped in a resonant cavity produce standing oscillations, whose temperature fluctuations are delayed by a quarter period with respect to the velocity signal, instead of the half-period delay from freely propagating waves \citep{Deubner1974, Felipe+etal2020}. In this study, we use numerical simulations and full-Stokes NLTE radiative transfer calculations to explore the generation of intensity emission cores in the \CaII\ 8542 \AA\ line during the downflowing stage of standing umbral oscillations, and how they can support the existence of the recently reported dowflowing UFs.

\section{Numerical methods}

Non-linear wave propagation between the solar interior and corona was simulated using the code MANCHA \citep{Khomenko+Collados2006, Felipe+etal2010a} in a two-dimensional domain. The details of the simulations can be found in Appendix \ref{appendix:numerical_setup}. We have computed synthetic full-Stokes spectra in the \CaII\ 8542 \AA\ line taking snapshots from our numerical simulation and then computing the spectra with the NICOLE code \citep[][see details in Appendix \ref{appendix:synthesis}]{SocasNavarro+etal2015}. The code assumes statistical equilibrium. \citet{ WedemeyerBohm+Carlsson2011} showed that non-equilibrium effects in \CaII\ are small at the formation height of the infrared triplet lines and, therefore, statistical equilibrium should suffice to model \CaII\ atoms. We calculated the transition probabilities between levels for \CaII\ in a sunspot model. They indicate characteristic transition times much shorter than one second (Appendix \ref{appendix:synthesis}), meaning that departures from statistical equilibrium are insignificant for the dynamics here studied.

\begin{figure}[!ht] 
 \centering
 \includegraphics[width=9cm]{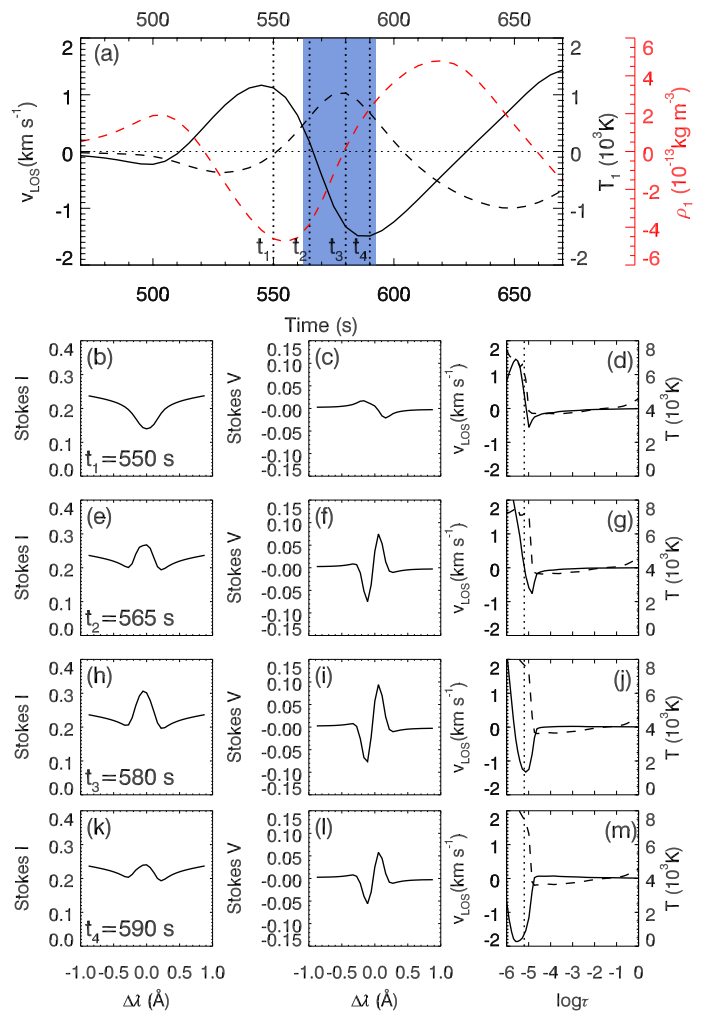}
  \caption{Development of an umbral flash produced by propagating waves as seen in \CaII\ 8542 \AA. Panel a: Temporal evolution of the vertical velocity (solid line, left axis), temperature (black dashed line, black right axis), and density (red dashed line, red right axis) fluctuations at $z=1.35$ Mm. Blue shaded region corresponds to the time when an umbral flash is identified in the Stokes profiles and the atmosphere is upflowing at $\log\tau=-5.1$. Lower panels: Stokes I (left column), Stokes V (middle column), and atmospheric models (right column). Each row corresponds to a different time step, as indicated at the bottom of the first column panels and the vertical dotted lines in panel a. In the right column, the vertical velocity (solid line, left axis) and temperature (dashed line, right axis) of the atmospheric model is plotted. The dashed vertical line marks the height of $\log\tau=-5.1.$}      
  \label{fig:UFpropagating}
\end{figure}

\begin{figure}[!ht] 
 \centering
 \includegraphics[width=9cm]{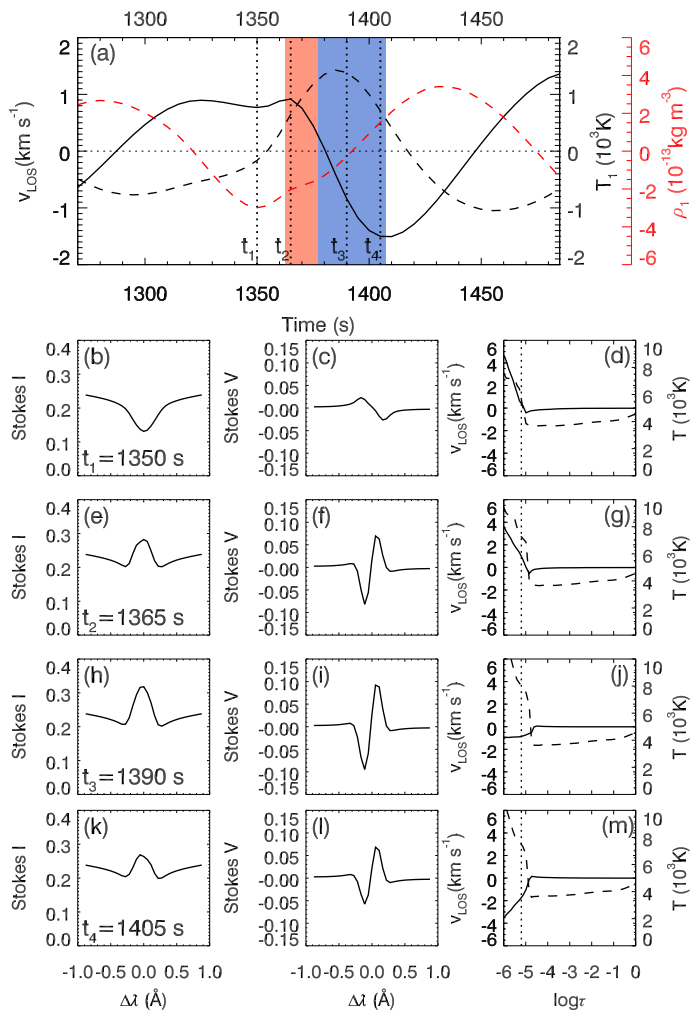}
  \caption{Same as Fig. \ref{fig:UFpropagating}, but for standing oscillations. Red shaded region in panel a indicates the time when an umbral flash is identified in the Stokes profiles and the atmosphere is downflowing at $\log\tau=-5.1.$}      
  \label{fig:UFstanding}
\end{figure}

\section{Results}
We analyze how UFs are generated in the case of propagating and standing oscillations. In Appendix \ref{appendix:simulations_standing_propagating}, we discuss the differences in the phase relations between temperature and velocity fluctuations from propagating and standing waves as determined from the analysis of numerical simulations. In the following, for the examination of the propagating-wave case, we focus on the first wavefront of the simulation reaching the chromosphere, before any previous wavefronts are reflected and a standing wave is settled. We chose to follow this approach (instead of using an independent simulation without transition region) in order to employ exactly the same atmospheric stratification for both cases. This way, we can assure that the differences in the properties of the UFs are produced by the nature of the oscillations and not by the changes in the background atmospheres.

\subsection{Umbral flashes from propagating waves}

Figure \ref{fig:UFpropagating} illustrates the development of UFs generated by a propagating wave at the beginning of the simulation. This case is qualitatively comparable to the synthetic UFs analyzed by \citet{Bard+Carlsson2010}, \citet{Felipe+etal2014b}, \citet{Felipe+etal2018a}, and \citet{Felipe+EstebanPozuelo2019}. We refer the reader to those papers for a detailed discussion of the formation of UFs. In the present paper, we focus on evaluating the chromospheric plasma motions during the appearance of the UF. Similarly to \citet{Henriques+etal2017}, we define a profile as flashed when the intensity around the core of the \CaII\ 8542 \AA\ line is above the value of Stokes I at $\Delta\lambda=0.942$ \AA. 

The time steps when flashed profiles take place are marked in blue in Figure \ref{fig:UFpropagating}a. As expected, they coincide with a temperature enhancement at $z=1.35$ Mm, which is roughly the height where the response of the \CaII\ 8542 \AA\ line to temperature is maximum during UFs (Fig. \ref{fig:RF}). The temperature maximum precedes the highest value of the upflow velocity by 9~s. Prior to the appearance of the flash (Fig. \ref{fig:UFpropagating}b) the atmosphere is downflowing agreeing with observations \citep{Henriques+etal2020}, but the temperature fluctuation is too low to produce an emission core (panel d). When the UF starts to develop (panel e) the atmosphere is upflowing at the optical depth where the response of the line is maximum  ($\log\tau\approx -5.1$, Fig. \ref{fig:RF}) and downflowing at higher layers (panel g). A similar situation can be seen at later steps (panels h and j). The change from upflows to downflows at a certain height is a manifestation of the propagating character of the waves since for standing waves we expect most of the chromosphere to be oscillating in phase.

\subsection{Umbral flashes from standing waves}
\label{sect:UFstanding}
The atmospheres generating UFs in the case of standing oscillations exhibit several significant differences from those from propagating waves. First, as discussed in Appendix \ref{appendix:simulations_standing_propagating}, the phase shift between velocity and temperature fluctuations is different. In the UF illustrated in Fig. \ref{fig:UFstanding}, the peak of the temperature enhancement at $z=1.35$~Mm takes place 26~s before the maximum value of the upflow velocity (panel a). Second, higher chromospheric layers are oscillating in phase above the resonant node as a result of the standing oscillations (right column from the bottom subset of Fig. \ref{fig:UFstanding}).

Due to the earlier appearance of a positive temperature fluctuation (in comparison with the case of propagating waves), they partially coincide with the downflow of the velocity oscillation. This way, chromospheric temperatures high enough to generate UFs are reached while a strong downflow is still taking place (panels e, f, and g from Fig. \ref{fig:UFstanding}). The strongest UF (at the peak of the temperature fluctuation) is found when the chromosphere is already upflowing (panels h, i, and j from Fig. \ref{fig:UFstanding}). In this example, the total duration of the UF is 45 s. During most of this time, the atmosphere at the height where the response of the \CaII\ 8542 line is maximum is upflowing. However, in the first 12.5 s of the flash, the atmosphere exhibits a downflow at $\log\tau=-5.1$ (red shaded region in Fig. \ref{fig:UFstanding}). The phase difference between velocity and temperature is not strictly $\pi/2$ (see Figs. \ref{fig:UFstanding}a and \ref{fig:simulations}c), but the temperature peak is slightly shifted towards the time when the atmosphere is upflowing, contributing to the longer duration of the upflowing stage of the UF. At the beginning of the UF ($t_2$ in Fig. \ref{fig:UFstanding}), the chromospheric density is reduced  (red dashed line in Fig. \ref{fig:UFstanding}a). The UF develops during the compressional phase of the oscillations and, thus, upflowing UFs are associated with a higher density.

\begin{figure}[!ht] 
 \centering
 \includegraphics[width=9cm]{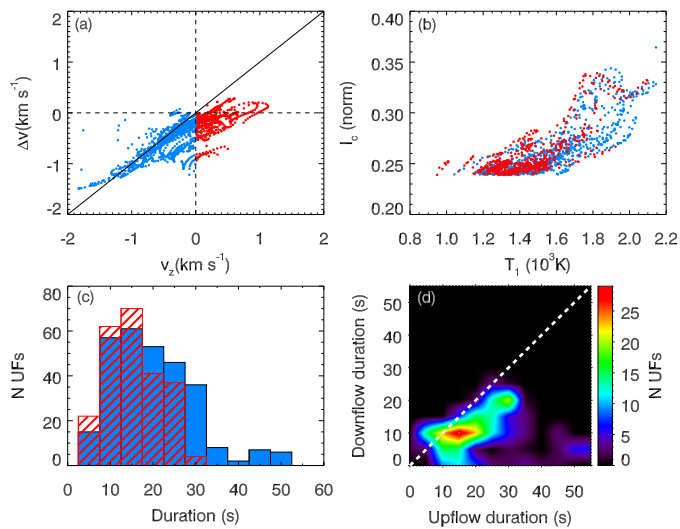}
  \caption{Properties of UFs in a simulation with a chromospheric resonant cavity. Panel a: Wavelength shift of the center of the emission core ($\Delta$v) as a function of the vertical velocity at $\log\tau=-5.1$ (v$_{\rm z}$). Panel b: Maximum intensity of the emission feature as a function of the temperature fluctuation at $\log\tau=-5.1$. In panels a and b, blue dots represent upflowing UFs and red dots correspond to downflowing UFs. Panel c: Histogram illustrating the duration of upflowing (blue bars) and downflowing (bars filled with red lines) UF events. Panel d: Relation between the duration of the upflowing UFs and the preceding downflowing UFs. The color bar indicates the number of occurrences of UFs with the corresponding durations.}      
  \label{fig:statistics}
\end{figure}

\subsection{Properties of standing waves umbral flashes}
\label{sect:UFstatistics}

We have identified a total of 1777 flashed profiles in the numerical simulation once the standing oscillations are established (see Appendix \ref{appendix:numerical_setup}). Approximately 38\% of those profiles are associated with downflows at $\log\tau=-5.1$, whereas the rest of the cases are produced during upflows. 

Figure \ref{fig:statistics} illustrates some of the properties of those UFs. Panel a shows the relation between the wavelength shift of the emission feature and the actual chromospheric velocity. For most of the UFs, the emission core is blueshifted, even when the atmosphere generating the flash is downflowing (see Fig. \ref{fig:UFstanding}e). \citet{Henriques+etal2017} showed how a downflowing UF can lead to an apparently blueshifted \CaII\ 8542 \AA\ emission core. Our analysis is based on the evaluation of the velocity at a fixed $\log\tau=-5.1$. We cannot discard changes in the velocity response function during UFs.


Figure \ref{fig:statistics}b shows the peak value of the emission core as a function of the temperature fluctuation. The latter is defined as the difference between the actual temperature and that from the background model (at rest) at $\log\tau=-5.1$ (6100 K). Chromospheric temperature enhancements of at least 1000~K are required to produce a flashed profile. The intensity of the core shows small variations for temperature perturbations up to 1600~K. Higher temperature enhancements generate the strongest UFs found in the simulation, exhibiting a significant increase in the intensity of the core. 

As previously pointed out, even in the case of standing oscillations most of the flashes are associated with upflows. Panels c and d from Fig. \ref{fig:statistics} examine the duration of the upflowing and downflowing stages of UF events. We define the duration as the temporal span while the intensity core at a certain spatial location is continuously in emission (above the same threshold previously defined). Since upflowing UFs are generally preceded by downflowing UFs (Sect. \ref{sect:UFstanding}), we have measured independently the duration of each of these phases. Figure \ref{fig:statistics}d shows that in the vast majority of the cases, the upflowing stage lasts longer than the downflowing phase of the same event (white dashed line marks the locations in the diagram where upflowing and downflowing UFs have the same duration). They are similar to the case illustrated in Fig. \ref{fig:UFstanding}. For both downflowing and upflowing phases the duration is generally in the range between 10 to 20 s, but the upflowing part of the UFs can extend longer than 25~s, whereas almost no downflows last that long. This explains the prevalence of detected UFs associated with upflows.

\section{Discussion and conclusions}
\label{sect:conclusions}

Umbral flashes are a well-known phenomenon and their relation to the oscillatory processes in sunspot atmospheres is an established fact. Early modeling of UFs has characterized them as the result of upward-propagating waves \citep{Havnes1970, Bard+Carlsson2010, Felipe+etal2014b}. In this scenario, most of the line formation of UFs takes place at layers where the chromospheric flows generated by the oscillation are upflowing. First spectropolarimetric analyses of UFs apparently confirmed this model \citep{SocasNavarro+etal2000a, delaCruz-Rodriguez+etal2013}. However, recent studies have showed indications of a chromospheric resonant cavity above sunspot umbrae \citep{Jess+etal2020a}, whose presence has been proved by \citet{Felipe+etal2020} based on the analysis of the phase relations of chromospheric oscillations. Umbral chromospheric waves do not propagate but form standing waves instead.

In this study, we have modeled the development of UFs produced by standing oscillations for the first time. The phase shift between temperature and velocity fluctuations from standing waves differs from that of purely propagating waves. In the latter, temperature enhancements are associated with upflows, leading to upflowing UFs, whereas in standing oscillations the temperature increase takes place a quarter period earlier. This way, standing waves generate UFs partially during the downflowing stage of the oscillation. Interestingly, recent studies have started to report downflowing UFs \citep{Henriques+etal2017, Houston+etal2020, Bose+etal2019}. \citet{Henriques+etal2017} suggested that dowflowing UFs may be related to the presence of coronal loops rooted in the sunspot umbra, since previous works have detected supersonic downflows in the transition region of such structures \citep[\eg,][]{Kleint+etal2014,Kwak+etal2016, Chitta+etal2016}. Our model offers an alternative explanation, interpreting downflowing UFs as a natural feature of the global process of sunspot oscillations. This is also the recent interpretation of \cite{Henriques+etal2020} but here we show that such downflows are evidence of the standing-wave nature of these oscillations. \citet{Henriques+etal2017} reported strong downflowing atmospheres for UFs, with velocities up to 10 km s$^{-1}$, whereas our model produces downflowing UFs with velocities around 1 km s$^{-1}$.


The development of UFs generated by standing waves is as follows. The chromospheric temperature starts to increase during the stage when the chromosphere is downflowing. All the chromospheric layers above the velocity resonant node are in phase and, thus, downflowing. This temperature increase is high enough to produce an emission core in the \CaII\ 8542 \AA\ line. During the peak of the temperature fluctuation, the chromospheric plasma changes from downflowing to upflowing. The profiles remain flashed for the upflowing atmospheres until the chromospheric temperature reaches again a value closer to that of a quiescent umbra. Unfortunately, the temporal cadence of most \CaII\ 8542 \AA\ observations up to date, in the range between 25-28~s \citep[\eg,][]{Henriques+etal2017,Joshi+delaCruzRodriguez2018} and 48~s \citep{Houston+etal2020}, is not sufficient to capture the evolution of UF events. The analysis is also challenged by the sudden variations of the atmosphere during the time required by Fabry-Perot spectrographs, such as the Interferometric BIdimensional Spectrometer \citep[IBIS,][]{Cavallini2006} and the CRisp Imaging SpectroPolarimeter \citep[CRISP,][]{Scharmer+etal2008}, to scan the line \citep{Felipe+etal2018b}. However, observations of small-scale umbral-brightenings using a cadence of 14~s exhibit some indication of a temporal evolution consistent with that predicted by our modeling \citep{Henriques+etal2020}. Upcoming observations with the National Science Foundation's Daniel K. Inouye Solar Telescope \citep[DKIST,][]{Rimmele+etal2020} will provide the ideal data for studying the temporal variations of UFs.

Our simulations show that the upflowing stage of UFs is generally longer than the downflowing part (Fig. \ref{fig:statistics}d). This is in agreement with the prevalence of upflowing solutions in NLTE inversions of observed UFs \citep{SocasNavarro+etal2000a, delaCruz-Rodriguez+etal2013}, even in studies where downflowing UFs are also reported \citep{Houston+etal2020}.

This letter provides a missing piece of the puzzle in what has been a convergence of observations and simulations by tying together the evidence for downflowing UFs and the existence of cavities in the chromosphere of the umbra as the latter will naturally generate the former. Finally, the case for a corrugated surface where downflows are significant \citep{Henriques+etal2020}, together with the importance of cavities for such downflows, implies that future observations constraining the transition region of sunspots should find the transition region itself to be highly corrugated. 







\begin{acknowledgements} 
Financial support from the State Research Agency (AEI) of the Spanish Ministry of Science, Innovation and Universities (MCIU) and the European Regional Development Fund (FEDER) under grant with reference PGC2018-097611-A-I00 is gratefully acknowledged. VMJH is funded by the European Research Council (ERC) under the European Union’s Horizon 2020 research and innovation programme (SolarALMA, grant agreement No. 682462) and by the Research Council of Norway through its Centres of Excellence scheme (project 262622). JdlCR is supported by grants from the Swedish Research Council (2015-03994) and the Swedish National Space Agency (128/15). This project has received funding from
the European Research Council (ERC) under the European Union's Horizon 2020 research and innovation program (SUNMAG, grant agreement 759548). The authors wish to acknowledge the contribution of Teide High-Performance Computing facilities to the results of this research. TeideHPC facilities are provided by the Instituto Tecnol\'ogico y de Energ\'ias Renovables (ITER, SA). URL: \url{http://teidehpc.iter.es}.
\end{acknowledgements}

\bibliographystyle{aa} 
\bibliography{biblio.bib}

\begin{appendix}

\section{Numerical setup}
\label{appendix:numerical_setup}

The ideal magnetohydrodynamic (MHD) equations were solved using the code MANCHA \citep{Khomenko+Collados2006, Felipe+etal2010a}. The numerical simulations were computed using the 2.5D approximation. We employed a two-dimensional domain, but vectors keep three-dimensional coordinates.  code solves the non-linear equations since UFs are characterized by acoustic shock waves. The shock capturing capabilities of the code have been validated by comparing its performance with analytic solutions of shock tube tests in both the MHD approximation \citep{Felipe+etal2010a} and including the effects of partial ionization \citep{GonzalezMorales+etal2018}. In this paper, we have restricted the analysis to ideal MHD. The vertical direction of the computational domain spans from $z=-1140$ km to $z=3500$ km, with $z=0$ set at the height where the optical depth at 5000 \AA\ is unity ($\log\tau=0$) and using a constant spatial step of 10 km. At top and bottom boundaries we imposed a perfectly matched layer \citep{Berenger1994}, which damps waves with minimum reflection. The horizontal domain includes 96 points with a spatial step of 50 km. Periodic boundary conditions were imposed in the horizontal direction.

The code computes the temporal evolution of waves driven below the solar surface as they propagate in a background umbral model. The background atmosphere is constant in the horizontal direction. The vertical stratification of a modified \citep{Avrett1981} model was established at all horizontal locations. The modifications of this umbral atmosphere include the extension of the model to deeper layers by smoothly merging it with model S \citep{Christensen-Dalsgaard+etal1996} and the shift of the chromospheric temperature increase to higher layers. The later was introduced to employ a model that generates \CaII\ 8542 \AA\ profiles comparable to those observed in actual observations, similar to \citet{Felipe+etal2018a}. The model includes a sharp temperature gradient from chromospheric temperature values to coronal temperatures starting at $z\approx1.8$ Mm. The atmosphere is permeated by a vertical magnetic field with constant strength of 2000 G. Waves are driven with a vertical force imposed at $z=-180$ km. The temporal and spatial dependence of the driver were extracted from temporal series measured in the photospheric \SiI\ 10827 \AA\ line \citep{Felipe+etal2018b} following the methods described in \citet{Felipe+etal2011}.

A total of 55 min of solar time were computed. The simulated atmospheres were saved with a temporal cadence of 5~s. For the analysis of the properties of UFs produced by standing waves (Sect. \ref{sect:UFstatistics}) we have avoided the first 15 min of the simulation since during this time the first wavefronts are arriving in the chromosphere and the standing oscillations are not established. Thus, the analysis is restricted to the last 40 min of simulations.

\begin{table}
\begin{center}
\caption[]{\label{table:transition_probabilities}
          {Transition probabilities per second at $\log\tau=-5$ in Maltby M sunspot model \citep{Maltby+etal1986}}}
\begin{tabular}{ccc}
\hline\noalign{\smallskip}
Transition	& P$_{3k}$ (s$^{-1}$)	        &   P$_{k3}$ (s$^{-1}$) 	\\
\hline\noalign{\smallskip}
1           & 2669	        &  19407 \\
2           & 73605	        &  49744               \\
4           & 23606          &  596             \\
5           & 1807920        &  26141             \\
6           & 4.10796$\times 10^{-2}$   &  1.47390$\times 10^{-1}$             \\
\hline

\end{tabular}

\begin{tablenotes}
\small
\item {Middle column indicates the transition probabilities from level 3 (\CaII\ 8542 \AA\ lower level) to all other levels $k$ (given by the first column). Right column shows the transition probabilities from level $k$ to level 3. Level $k=6$ is the \CaII\ continuum.}
\end{tablenotes}

\end{center}
\end{table}

\begin{figure}[!ht] 
 \centering
 \includegraphics[width=9cm]{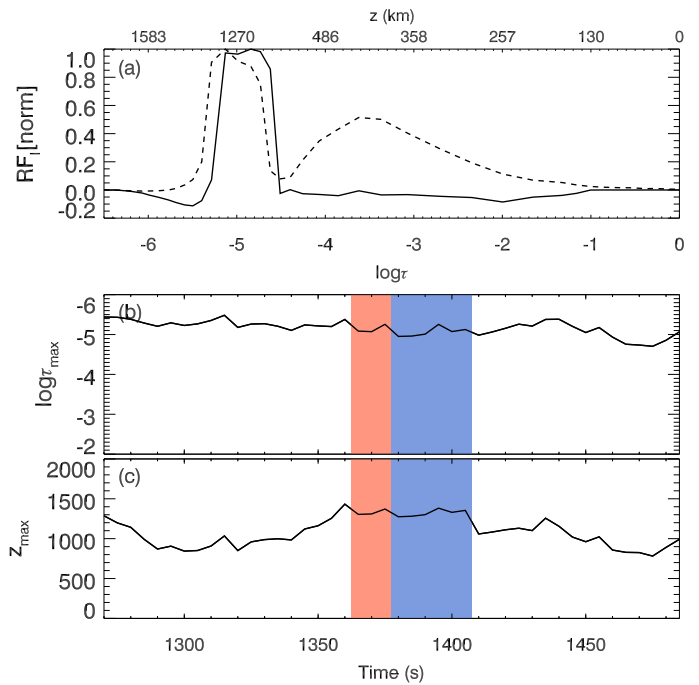}
  \caption{Response functions of the intensity profile of the \CaII\ 8542 \AA\ line during the development of an UF. Panel a: Response function of Stokes I to the temperature at the core of the line (dashed line) and to the velocity at the wavelength where the response is maximum (8542.2 \AA, solid line) at $t=1405$ s (see profile in Fig. \ref{fig:UFstanding}k). Lower axis corresponds to the optical depth and upper axis to the geometrical height. Panel b: Optical depth of the maximum of the response function to temperature. Panel c: Geometrical height of the maximum of the response function to temperature. Color-shaded regions in panels b and c have the same meaning as in Fig. \ref{fig:UFstanding}.}  
  \label{fig:RF}
\end{figure}

\begin{figure}[!ht] 
 \centering
 \includegraphics[width=9cm]{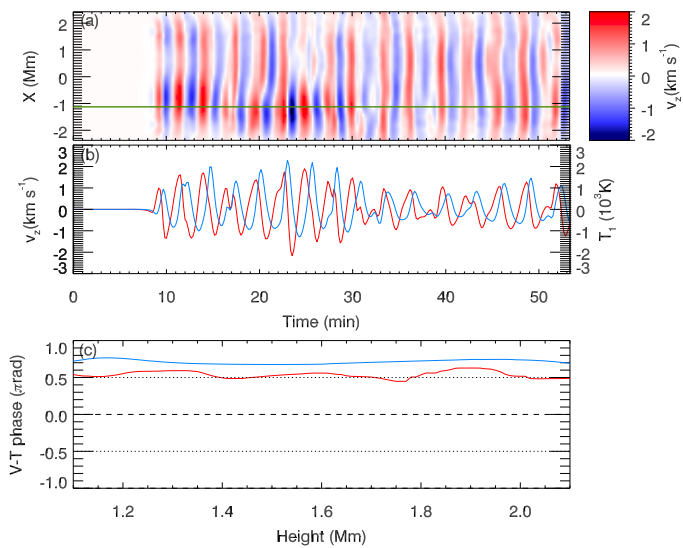}
  \caption{Numerical simulations of wave propagation in a sunspot umbra. Panel a: Vertical velocity at $z=1.35$ Mm for all the horizontal positions. Panel b: Vertical velocity (red line, left axis) and temperature fluctuations (blue line, right axis) at $X=-1.1$ Mm (location indicated by the green vertical line in panel a). Panel c: Average phase difference between velocity and temperature oscillations for waves with a frequency between 5 and 6 mHz as a function of atmospheric height. The red line represents the phase shift for standing waves from a simulation with a resonant cavity. The blue line corresponds to propagating waves from a simulation without a transition region. Dotted (dashed) lines indicate a phase shift of $\pm\pi/2$ (0).}      
  \label{fig:simulations}
\end{figure}

\section{Spectral synthesis}
\label{appendix:synthesis}

The Stokes profiles of the \CaII\ 8542 \AA\ line were synthesized for all the spatial locations and time steps of the simulation using the NLTE code NICOLE \citep{SocasNavarro+etal2015}. NICOLE was fed with the velocity, temperature, gas pressure, density, and magnetic field vector stratification of each atmosphere (for a specific horizontal position and time) in a geometrical scale, as given by the output of the numerical simulation. An artificial macroturbulence of 1.8 km s$^{-1}$ was added to balance the absence of small scale motions in the numerical calculations and generate broader line profiles, similar to those measured in actual observations. The Stokes profiles were first synthesized with a spectral resolution of 5 m\AA. They were then degraded to a resolution of 55 m\AA\ by convolving them with a 100 m\AA\ full width half maximum Gaussian filter centered at the wavelengths chosen for the spectral sampling.

Table \ref{table:transition_probabilities} shows the transition probabilities from the lower level of the \CaII\ 8542 \AA\ line to all other levels and vice versa. The large values point to very short characteristic transition times (below one second), confirming that the assumption of statistical equilibrium leads to very small decay times, shorter than the temporal scales of interest for our study. The only exception are the rates to and from the \CaII\ continuum ($k=6$). The characteristic times for this transition is of the order of minutes, which means that ionization to \CaIII\ might
be lagging behind the dynamics. However, this is not relevant for our purposes since \CaII\ is the dominant species.

Figure \ref{fig:RF} illustrates the response function of the \CaII\ 8542 \AA\ intensity to the temperature during the development of an UF. The main contribution to Stokes I comes from an optical depth around $\log\tau=-5.1$. We have chosen this optical depth as the reference for the analyses presented in this study since it is maintained as the main contribution to the formation of the line during the whole temporal span of UFs (color-shaded regions in Fig. \ref{fig:RF}b). During this time, $\log\tau=-5.1$ samples a geometrical height around $z=1350$ km (Fig. \ref{fig:RF}c). Remarkable changes in the chromospheric region with higher sensitivity to the temperature are found between flashed and non-flashed profiles (ranging between $\log\tau=-4.7$ and $\log\tau=-5.6$).

\section{Standing and propagating oscillations}
\label{appendix:simulations_standing_propagating}

Figure \ref{fig:simulations} illustrates the chromospheric oscillations from the numerical simulation. Panel a shows the velocity wavefield for all the umbral spatial positions included in the simulation. We use the usual spectroscopic velocity convention in astrophysics where negative velocities correspond to upflows (blueshifts, indicated in blue in Fig. \ref{fig:simulations}a) and positive velocities represent downflows (redshifts, red in Fig. \ref{fig:simulations}a). Figure \ref{fig:simulations}b exhibits the temporal evolution of velocity (red line) and temperature (blue line) fluctuations at a randomly selected umbral location. Temperature wavefronts are delayed with respect to velocity wavefronts. For a better examination, Fig. \ref{fig:simulations}c shows the phase difference between temperature and velocity oscillations for waves with frequencies in the range 5-6 mHz (red line). Only the atmospheric heights of interest for the formation of the \CaII\ infrared lines are plotted. The phase shift is around $\pi/2$, indicating that temperature fluctuations lag velocity oscillations by a quarter period, in agreement with theoretical estimates of standing waves \citep{Deubner1974, Al+etal1998}.

The blue line in Fig. \ref{fig:simulations}c corresponds to the phase-shift derived from a numerical simulation where the transition region is removed. Instead, a constant temperature atmosphere is set above $z=1.5$ Mm \citep{Felipe2019, Felipe+etal2020}. In this situation, waves can freely propagate in the upper atmospheric layers and the phase relation between velocity and temperature fluctuations tends to $\pi$, the expected value for propagating waves. The phase-shift from purely propagating waves (strictly $\pi$) is not obtained due to two reasons. First, there are temperature gradients in the model (the temperature increase from the photospheric to the chromospheric temperature), which produce some partial wave reflections at $z\approx1.6$ Mm. Second, as waves propagate upwards they develop into shocks and temperature enhancements take place earlier, shifting towards the compression phase of the wave. Summarizing, in the propagating-waves scenario (blue line in Fig. \ref{fig:simulations}c) temperature fluctuations are approximately in anti-phase with velocity oscillations, with temperature enhancements corresponding to upflows. If a chromospheric resonant cavity is present (red line), temperature oscillations are shifted backward and their increase partially coincides with the downflowing stage of the oscillations. For waves with $\approx$5.5 mHz frequency, from $z=1.1$ Mm to $z=1.8$ Mm the shift of temperature fluctuations is in the range between 25 and 80~s.

\end{appendix}

\end{document}